%% file: main.tex
\documentclass[letterpaper]{article} % DO NOT CHANGE THIS
\usepackage{aaai25}  % DO NOT CHANGE THIS
\usepackage{times}  % DO NOT CHANGE THIS
\usepackage{helvet}  % DO NOT CHANGE THIS
\usepackage{courier}  % DO NOT CHANGE THIS
\usepackage[hyphens]{url}  % DO NOT CHANGE THIS
\usepackage{graphicx} % DO NOT CHANGE THIS
\usepackage{natbib}  % DO NOT CHANGE THIS AND DO NOT ADD ANY OPTIONS TO IT
\usepackage{caption} % DO NOT CHANGE THIS AND DO NOT ADD ANY OPTIONS TO IT
\usepackage{algorithm}
\usepackage{newfloat}
\usepackage{listings}
\DeclareCaptionStyle{ruled}{labelfont=normalfont,labelsep=colon,strut=off} % DO NOT CHANGE THIS
\floatstyle{ruled}
\newfloat{listing}{tb}{lst}{}
\floatname{listing}{Listing}
\input{header}

\begin{document}

%\title{A Multi-stage Anti-blur Network for Unsupervised Deformable Image Registration}

%\title{UniBrain: A Unified Model for End-to-End Brain Imaging Tasks}
\title{End-to-End Deep Learning for Structural Brain Imaging: A Unified Framework}
\author{
    Yao Su\textsuperscript{\rm 1},
    Keqi Han\textsuperscript{\rm 2},
    Mingjie Zeng\textsuperscript{\rm 1},
    Lichao Sun\textsuperscript{\rm 3},
    Liang Zhan\textsuperscript{\rm 4},
    Carl Yang\textsuperscript{\rm 2}, \\
    Lifang He\textsuperscript{\rm 3},
    Xiangnan Kong\textsuperscript{\rm 1}
}
\affiliations{
    \textsuperscript{\rm 1}Worcester Polytechnic Institute, Worcester, MA, USA\\
    \textsuperscript{\rm 2}Emory University, Atlanta, GA, USA\\
    \textsuperscript{\rm 3}Lehigh University, Bethlehem, PA, USA\\
    \textsuperscript{\rm 4}University of Pittsburgh, Pittsburgh, PA, USA\\
    ysu6@wpi.edu, keqi.han@emory.edu, mzeng2@wpi.edu, lis221@lehigh.edu, liang.zhan@pitt.edu, j.carlyang@emory.edu, lih319@lehigh.edu, xkong@wpi.edu
}

\maketitle

\input{sec_0_abstract_v3}





% Introduction Section
% \input{sec_1_intro_v3}
\input{sec_1_intro}

% %-----------------------------------------------
% % Related Work Section
%\input{sec_5_related_work}

% %-----------------------------------------------
% % Problem Definition Section
%\input{sec_2_problem_def3}

% %-----------------------------------------------
% % Method Section
\input{sec_3_method}

% %-----------------------------------------------
% % Experiment Section
\input{sec_4_experiment}

% %-----------------------------------------------
% % Conclusion Section
\input{sec_6_conclusion}
%\newpage

% balance the two columns of the last page
%\balance

% \bibliographystyle{ACM-Reference-Format}
\bibliography{header,reference, ref_he}
\end{document}

%% file: header.tex
\usepackage{amsthm}
\usepackage{algpseudocode}
\usepackage{amsmath}
\usepackage{amssymb}
\usepackage{bm}
\usepackage{color}
\usepackage{etex}
\usepackage{qtree}
\usepackage{graphicx}
\usepackage{multirow}
\usepackage{multicol}
\usepackage{subfigure}
\usepackage{url}
\usepackage{thmtools}
\usepackage{ctable}
\usepackage{tabularx}
\usepackage{booktabs}   % professional tables. See http://cs.brown.edu/about/system/managed/latex/doc/booktabs.pdf
\usepackage{graphics}   % image files in figure (e.g., pdf, eps files)
\usepackage{url}        % formating a web url using \url{...}
\usepackage{pifont}     % adding special characters using \ding{...}. See http://willbenton.com/wb-images/pifont.pdf
\usepackage{xcolor}      % color text
\usepackage{xspace}
\newcommand{\ie}[0]{\textit{i.e.},\ }   % i.e., meaning "which is/means "
\newcommand{\eg}[0]{\textit{e.g.},\ }   % e.g., meaning "for example, "
\usepackage[labelfont={bf,small},textfont={small}]{caption}
\usepackage{pifont}% http://ctan.org/pkg/pifont
\newcommand{\cmark}{\ding{51}}%
\newcommand{\xmark}{\ding{55}}%

\newcommand{\eat}[1]{}

%% file: sec_0_abstract_v3.tex
\begin{abstract}
Brain imaging analysis is fundamental in neuroscience, providing valuable insights into brain structure and function. 
Traditional workflows follow a sequential pipeline—brain extraction, registration, segmentation, parcellation, network generation, and classification—treating each step as an independent task. These methods rely heavily on task-specific training data and expert intervention to correct intermediate errors, making them particularly burdensome for high-dimensional neuroimaging data, where annotations and quality control are costly and time-consuming. 
We introduce UniBrain, a unified end-to-end framework that integrates all processing steps into a single optimization process, allowing tasks to interact and refine each other. Unlike traditional approaches that require extensive task-specific annotations, UniBrain operates with minimal supervision, leveraging only low-cost labels (\ie classification and extraction) and a single labeled atlas. By jointly optimizing extraction, registration, segmentation, parcellation, network generation, and classification, UniBrain enhances both accuracy and computational efficiency while significantly reducing annotation effort. Experimental results demonstrate its superiority over existing methods across multiple tasks, offering a more scalable and reliable solution for neuroimaging analysis.

\end{abstract}

%% file: sec_1_intro.tex
\vspace{-5pt}
\section{Introduction}
\label{sec:intro}
\input{fig_problem}
The human brain, with its billions of interconnected neurons that form the connectome, is the foundation of our cognitive functions and behaviors. Understanding this intricate connectivity is crucial for decoding the brain's mechanisms in development and degeneration. However, accurately mapping the connectome remains a significant challenge due to limitations in current methods. Traditional workflows rely on structural or functional neuroimaging data processed through fragmented steps—brain extraction, registration, segmentation/parcellation, and network generation—often requiring manual quality control, which is costly and represents a critical barrier for quantitative brain biomarkers to enter clinical practice. Furthermore, piecemeal approaches prevent simultaneous optimization of interdependent stages, leading to inefficiencies and limiting the discovery of nuanced connections. Errors introduced in earlier steps propagate through subsequent analyses, resulting in potentially misleading interpretations of brain dynamics. Moreover, the time-intensive nature of these workflows hinders scalability and efficiency. 

Instead of tedious, step-by-step processing for brain imaging data, recent studies support transforming these pipelines into deep neural networks for joint learning and end-to-end optimization \cite{ren2024deepprep, agarwal2022end}. While several approaches have been proposed—such as joint extraction and registration \cite{su2022ernet}, joint registration and parcellation \cite{zhao2021deep, lord2007simultaneous}, and joint network generation and disease prediction \cite{campbell2022dbgsl, mahmood2021deep, kan2022fbnetgen}—there is currently no framework that unifies and simultaneously optimizes all these processing stages to directly create brain networks from raw imaging data. Mapping the connectome of human brain as a brain network (\ie graph), has become one of the most pervasive paradigms in neuroscience \cite{sporns2005human,bargmann2013connectome}. Representing the brain as a graph of nodes (regions) and edges (structural or functional connections) enables gaining critical insights into brain organization, identifying key regions or hubs, and understanding how brain connectivity changes under different conditions (\eg during development, aging, or neurological disorders) \cite{kaiser2011tutorial,crossley2014hubs,xu2015connectome}. This need has intensified with the rapidly advancing imaging technologies and massive data collection.

In this paper, we propose UniBrain, the first end-to-end deep learning model that seamlessly integrates brain extraction, registration, segmentation, parcellation, network generation, and clinical classification into a unified optimization process, as illustrated in Figure~\ref{fig:intro}. Our objective is to investigate the interdependence of these tasks, enabling them to enhance each other's performance while relying on minimal labeled data. Specifically, we leverage low-cost labels (\ie extraction mask, classification label) and a single labeled template (\emph{a.k.a.} atlas) to jointly optimize all tasks. Notably, our approach eliminates the need for instance-level ground-truth labels for registration, segmentation, parcellation, and network connectivity during model training. Extensive experiments on the public ADHD dataset with 3D brain sMRI demonstrate that our method outperforms state-of-the-art approaches across all six tasks.

\vspace{-5pt}
\section{Related Works}
\label{sec:related}
In the literature, related tasks in brain imaging analysis have been extensively studied. Conventional methods primarily focus on designing methods for brain extraction~\cite{kleesiek2016deep,lucena2019convolutional}, registration~\cite{sokooti2017nonrigid, su2022abn}, segmentation~\cite{akkus2017deep, kamnitsas2017efficient, chen2018voxresnet}, parcellation~\cite{thyreau2020learning,lim2022deepparcellation}, network generation~\cite{vskoch2022human, yin2023multi} and classification~\cite{li2021braingnn,kawahara2017brainnetcnn, kan2022brain} separately under supervised settings. However, in brain imaging studies, the collection of voxel-level annotations, transformations between images, and task-specific brain networks often prove to be expensive, as it demands extensive expertise, effort, and time to produce accurate labels, especially for high-dimensional neuroimaging data, \eg 3D MRI. To reduce this high demand for annotations, recent works have utilized automatic extraction tools~\cite{smith2002fast,cox1996afni,shattuck2002brainsuite, segonne2004hybrid}, unsupervised registration models~\cite{balakrishnan2018unsupervised,su2022abn}, inverse warping~\cite{jaderberg2015spatial}, and correlation-based metrics~\cite{liang2012effects} for performing extraction, registration, segmentation, parcellation and network generation. Nevertheless, these pipeline-based approaches frequently rely on manual quality control to correct intermediate results before performing subsequent tasks. Conducting such visual inspections is not only time-consuming and labor-intensive but also suffers from intra- and inter-rater variability, thereby impeding the overall efficiency and performance. More recently, joint extraction and registration~\cite{su2022ernet}, joint registration and segmentation~\cite{xu2019deepatlas}, joint extraction, registration and segmentation~\cite{su2023one}, and joint network generation and classification~\cite{kan2022fbnetgen} have been developed for collective learning. However, partial joint learning overlooks the potential interrelationships among these tasks, which can adversely affect overall performance and limit generalizability. There is a pressing need for more integrated, automated and robust methodologies that can seamlessly integrate and optimize all stages of raw brain imaging-to-graph analysis within a unified framework.

%% file: fig_problem.tex
\begin{figure}[t]
  \centering
  \includegraphics[width=1.0\linewidth]{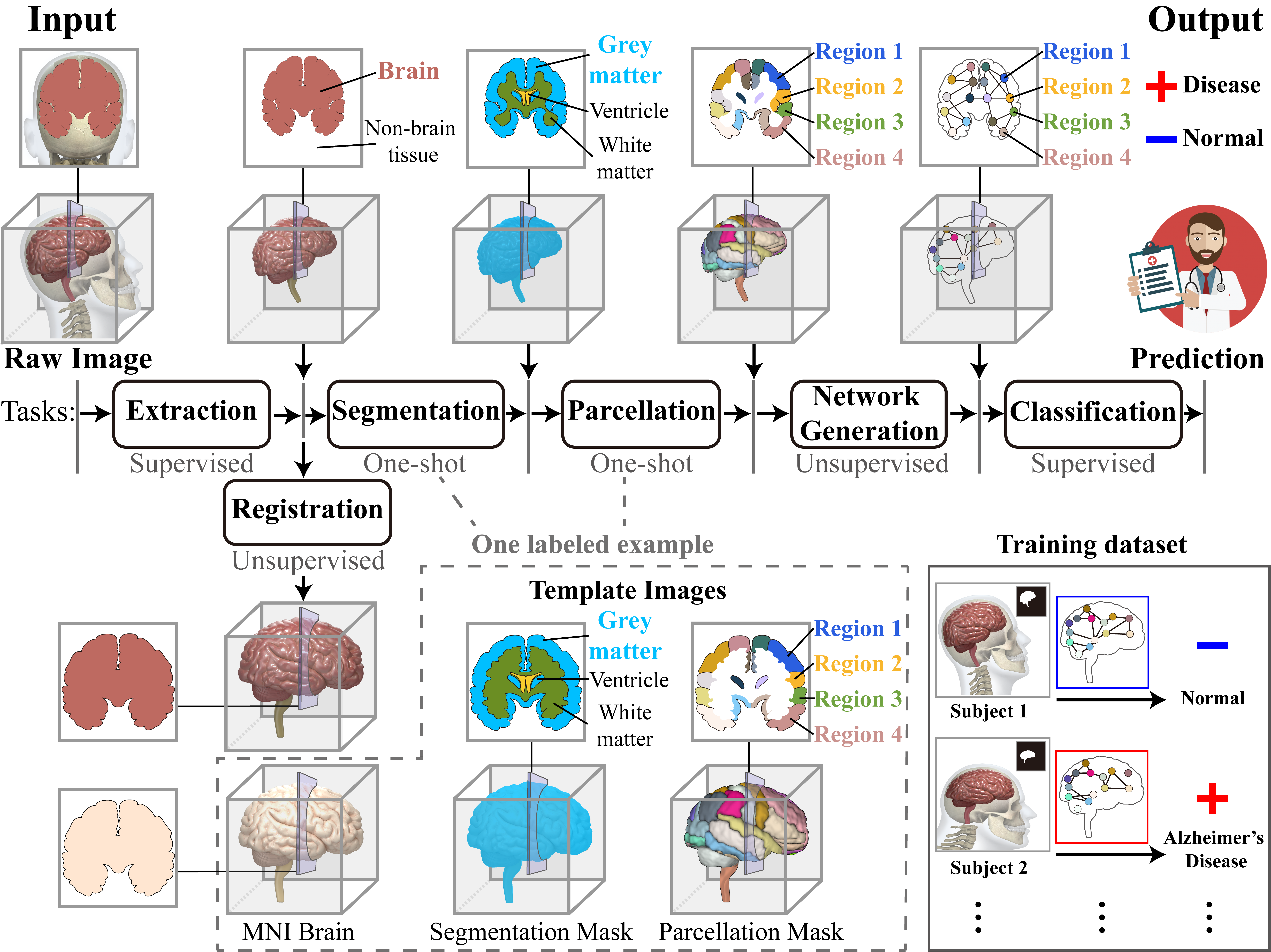}
  \vspace{-10pt}
  \caption{
The problem of end-to-end learning for brain imaging tasks. Given a set of raw images, each with a corresponding extraction mask and diagnosis label, along with a labeled template brain (with segmentation and parcellation masks), the goal is to train a model to simultaneously perform extraction, registration, segmentation, parcellation, network generation, and classification tasks.
  }
  \label{fig:intro}
  \vspace{-10pt}
\end{figure}

%% file: sec_3_method.tex
\vspace{-5pt}
\section{Our Approach}
\label{sec:method}
%\noindent\textbf{Overview.} 
%Figure~\ref{fig:network} presents the overview of the UniBrain framework. 
%designed for the end-to-end brain imaging analysis problem. 
%Our method is an end-to-end deep neural network consisting of five main modules: 1) \emph{Extraction Module} processes the raw source image $\mathbf{S}$ to yield the extracted brain image $\mathbf{E}$; 2) \emph{Registration Module} aligns extracted brain image $\mathbf{E}$ with the target image $\mathbf{T}$, resulting in the warped image $\mathbf{W}$; 3) \emph{Segmentation \& Parcellation Modules} leverage the target segmentation mask $\mathbf{B}$ and parcellation mask $\mathbf{P}$ to produce the source segmentation mask $\mathbf{R}$ and parcellation mask $\mathbf{U}$; 4) \emph{Brain Network Module} generate the brain network $G$ based one the region information provided by source image $\mathbf{S}$ and parcellation mask $\mathbf{U}$; 5) \emph{Classification Module} employs the generated brain network $G$ for prediction, outputting $\hat{y}$. The final output of UniBrain includes:  extracted brain image $\mathbf{E}$ constituting only cerebral tissues; warped image $\mathbf{W}$ aligning with the target image; brain segmentation mask $\mathbf{R}$ and parcellation mask $\mathbf{U}$ indicating tissue types and regions of the source image; brain network $G$ representing connectivity among regions, and prediction $\hat{y}$ signifying classification outcomes. 

UniBrain integrates multiple modules for brain extraction, registration, segmentation, parcellation, network generation, and classification, seamlessly connecting them within an end-to-end framework to enable collective learning.
Below, we provide a detailed description of each module.

\vspace{-3pt}
\subsection{Extraction Module}

The extraction module aims to extract brain from the raw image with assistance from two components:

\vspace{1pt}
\noindent \textbf{Extraction Network: $f_{e}$}. The extraction network $f_{e}(\cdot)$ acts as an annotator, intended to identify brain and non-brain tissues in the source image $\mathbf{S}$ and delineate their locations, thus providing the guidance for subsequent non-brain tissue elimination. Specifically, we employ the 3D U-Net as the base network to learn $f_{e}(\cdot)$.
The process can be formally expressed as:

\begin{equation}
\hat{\mathbf{M}}=f_{e}\left(\mathbf{S}\right),
\end{equation}
where $\hat{\mathbf{M}}$ is predicted extraction mask. During inference, $\hat{\mathbf{M}}$ is binarized by a Heaviside step function.

\vspace{1pt}
\noindent \textbf{Overlay Layer: $OL$}.
The overlay layer serves to eliminate non-brain tissues by applying the predicted brain mask $\hat{\mathbf{M}}$ to the source image $\mathbf{S}$. The final extracted image is $ \mathbf{E} = \mathbf{S}\circ \hat{\mathbf{M}}$, where $\circ$ denotes the element-wise multiplication.

\vspace{-3pt}
\subsection{Registration Module}
\label{sec: registration Module}
The registration module aims to align the extracted image with the target image, providing transformations for subsequent segmentation and parcellation tasks. This module comprises two main components:

\vspace{1pt}
\noindent \textbf{Registration Network: $f_{r}$}.
The registration network $f_{r}(\cdot, \cdot)$ processes the extracted image $ \mathbf{E}$ and target image $\mathbf{T}$ to learn the affine transformation $\mathbf{A}$, which establishes the coordinate correspondence between source and target image space. A 3D CNN-based encoder is used to learn $f_{r}(\cdot, \cdot)$ as:
\begin{equation}
\mathbf{A}=f_{r}\left(\mathbf{E}, \mathbf{T}\right).
\end{equation}
We leverage the multi-stage registration technique~\cite{su2022abn,zhao2019recursive} to boost registration performance, where $\mathbf{E}$ is recursively aligned with $\mathbf{T}$ though $M$ stages. 
%A study of $M$ can be found in Appendix A.2.1.

\vspace{1pt}
\noindent \textbf{Spatial Transformation Layer: $STL$}.
A key step in image registration is reconstructing the warped image $\mathbf{W}$ from the extracted image $\mathbf{E}$ using the affine transformation $\mathbf{A}$. This warping process is facilitated by a spatial transformation layer (STL), which resamples voxels from the extracted
image $\mathbf{E}$ to produce the warped image $\mathbf{W} = \mathcal{T}(\mathbf{E}, \mathbf{A})$. Given the affine transformation operator, we hold
\begin{equation}
     \mathbf{W}_{xyz} = \mathbf{E}_{x'y'z'} \hspace{1pt},
     \label{equ:voxel_value_k}
\end{equation}
where coordinate correspondence $[x', y', z', 1]^\top = \mathbf{A}[x, y, z, 1]^\top $. 
To enable successful gradient propagation, we use a differentiable transformation based on trilinear interpolation proposed by~\cite{jaderberg2015spatial}.

\vspace{-3pt}
\subsection{Segmentation \& Parcellation Module}
The segmentation and parcellation module creates segmentation and parcellation masks on the source image.
Leveraging recent developments in one-shot learning~\cite{wang2020lt, ding2021modeling, su2023one}, the module can generate these masks using a single labeled template image. The module contains two main components:

\vspace{1pt}
\noindent \textbf{Inverse Warping} Utilizing a single labeled example (\ie target image $\mathbf{T}$ with its corresponding segmentation mask $\mathbf{B}$ and parcellation mask $\mathbf{P}$) and the learned affine transformation $\mathbf{A}$, we apply the inverse transformation $\mathbf{A}^{-1}$ to generate warped segmentation mask $\mathbf{V} = \mathcal{T}(\mathbf{B}, \mathbf{A}^{-1})$ and parcellation mask $\mathbf{U} = \mathcal{T}(\mathbf{P}, \mathbf{A}^{-1})$ in the source image space as:
\begin{equation}
    \mathbf{V}_{cxyz} = \mathbf{B}_{cx'y'z'}, \forall  c \in \{1 ,\ldots, C\},
\end{equation}
\begin{equation}
    \mathbf{U}_{kxyz} = \mathbf{P}_{kx'y'z'}, \forall  k \in \{1 ,\ldots, K\},
\end{equation}
where coordinate correspondence $[x', y', z', 1]^\top = \mathbf{A}^{-1}[x, y, z, 1]^\top $, $c$ is the index for tissue class and $k$ is the index for ROIs. Same as the $\text{STL}$ layer in Registration Module, we then apply a differentiable transformation based on trilinear interpolation. 

\vspace{1pt}
\noindent \textbf{Segmentation Network: $f_s$}. The segmentation network $f_s(\cdot)$ aims to generate a segmentation mask for the source image $\mathbf{S}$ that 
matches the synthesized warped segmentation mask $\mathbf{V}$. We employ the widely-used 3D U-Net as the base network to learn $f_s(\cdot)$. Formally, we have:
\begin{equation}
    \mathbf{R} = f_s(\mathbf{S}).
\end{equation}

%In our study, we adopted different strategies for brain segmentation and parcellation: 1) For segmentation, neural networks are used to learn the source image's segmentation mask $\mathbf{R}$, effectively identifying brain tissues with clear boundaries and adjusts well to variations between tissues, resulting in precise segmentation masks. This process provide positive feedback to the registration module, enhancing label accuracy~\cite{wang2020lt, zhao2019data, ding2021modeling, su2023one}; 2) For parcellation, due to its complexity and less defined boundaries in functional regions, we used inverse warping to obtain the source's parcellation mask $\mathbf{U}$. Additionally, the choice of parcellation atlas can vary (\eg AAL and Desikan)~\cite{arslan2018human}.

\input{tab_res_ADHD.tex}

\input{tab_methods_v4}

\vspace{-3pt}
\subsection{Brain Network Module}

The brain network module generates the brain network using ROI information from parcellation mask $\mathbf{U}$ and the source image $\mathbf{S}$. The modules include three components:

%\subsubsection{Overlay Layer: $OL$} Similar to the operation in Section~\ref{sec: overlap}, this component is responsible for isolating each ROI image from the source image $\mathbf{S}$ using parcellation mask $\mathbf{U}$. First, $\mathbf{S}$ is expanded to $\tilde{\mathbf{S}} \in \mathbb{R}^{K \times W \times H \times D}$, where each \( k^{\text{th}} \) slice of \( \tilde{\mathbf{S}} \) is a copy of \( \mathbf{S} \). The ROI extracted image $\mathbf{F} \in \mathbb{R}^{K \times W \times H \times D} = \tilde{\mathbf{S}} \circ \mathbf{U}$ is then generated by applying an element-wise product~$\circ$ between $\tilde{\mathbf{S}}$ and $\mathbf{U}$.

\vspace{1pt}
\noindent \textbf{Overlay Layer: $OL$}. Similar to $OL$ in the Extraction Module, this component is responsible for isolating each ROI from the source image $\mathbf{S}$ using parcellation mask $\mathbf{U}$. The parcellated image $\mathbf{F} = \mathbf{S} \circ \mathbf{U}$ is generated by applying an element-wise product~$\circ$ between $\mathbf{S}$ and $\mathbf{U}$.

\vspace{1pt}
\noindent \textbf{Brain Network Function: $f_o$}. The brain network function aims to learn the representation for each ROI within the parcellated image $\mathbf{F}$. A weight-sharing Multilayer Perceptron (MLP) is employed to learn $f_o(\cdot)$, ensuring consistent feature extraction and generalization, which is expressed as:
\begin{equation}
    \mathbf{H}_{k} = f_o(\mathbf{F}_{k}), \forall  k \in \{1 ,\ldots, K\},
\end{equation}
where $k$ is the index for the ROIs. 
%The concatenated ROI feature $\mathbf{H}$ is then used to construct a task-aware brain network.

\vspace{1pt}
\noindent \textbf{Brain Network Generation}. The step generates a brain network based on the similarity between ROI representation pairs. Without loss of generality, here we use inner-product to measure the edge weights of the brain network. However, other differentiable similarity functions (\eg Mahalanobis distance and cosine similarity) can be used. To compute the connectivity matrix $\mathbf{C}$, each ROI representation $\mathbf{H}_{k}$ is first normalized with the $\ell^{2}\text{-norm}$, followed by the inner-product:
\begin{equation}
    \mathbf{C} = \mathbf{H} \mathbf{H^\top}.
\end{equation}
%the similarity between ROI representation pairs serves as edge weight of the network, expressed as: 
%The process begins by normalizing each ROI feature $\mathbf{H}_{k}$ with the $\ell^{2}\text{-norm}$, setting the stage to compute the connectivity matrix as: $\mathbf{C} = \mathbf{H} \mathbf{H^\top}$.
%This normalization ensures the stabilization of the learning process and maintains consistent weight magnitudes across the network, scaling the values of $\mathbf{C}$ to the range of $[-1, 1]$.
This normalization scales the values of $\mathbf{C}$ to the range of $[-1, 1]$, ensuring the stabilization of the learning process and maintaining consistent weight magnitudes in the network.
%The similarity score indicates the ROIs closer in embedding space are more likely to be connected, as suggested by~\cite{grover2019graphite, kipf2016variational,zou2019encoding,kan2022fbnetgen}.
%To further refine the network, connections in $\mathbf{C}$ with negative weights are screened out to reduces complexity and potential noise from less relevant connections~\cite{li2021braingnn, said2023neurograph, van2010comparing,kim2021learning}.

\subsection{Classification Module}
%The classification module makes the final diagnostic prediction. 
The classification module makes a final predictive diagnosis.

\vspace{1pt}
\noindent \textbf{Classification Network: $f_g$}. The classification network $f_{g}(\cdot, \cdot)$ aims to make a prediction based on the generated brain network while feeding task-specific insights to the preceding module, facilitating the brain network generation. We leverage the GCN~\cite{kipf2017semi} as the base network. The prediction $\hat{y}$ is obtained as:

\begin{equation}
\hat{y} = f_g(\mathbf{C},\mathbf{H}),
\end{equation}
where $\mathbf{H}$ is the initial node features and $\mathbf{C}$ is the learnable connectivity matrix provided by the brain network module.
%provided by the preceding brain network module.

\vspace{-3pt}
\subsection{End-to-End Training}
\label{section:end-to-end training}

We train UniBrain by minimizing the objective function as: 
\begin{equation}
\label{eq:optimization}
\begin{aligned}
\mathcal{L} = \hspace{+1pt} \mathcal{L}_{cls}\big(\hat{y}, y\big)  & + \alpha \mathcal{L}_{ext}\big(\hat{\mathbf{M}}, \mathbf{M}\big)   + \\ \beta  \mathcal{L}_{sim}\big(\mathbf{W}, \mathbf{T}\big)   & + \gamma \mathcal{L}_{seg}\big({\mathbf{R}}, \mathbf{V}\big),
\end{aligned}
\end{equation}
where $\mathcal{L}_{cls}(\cdot, \cdot)$ is classification loss term, $\mathcal{L}_{ext}(\cdot, \cdot)$ is extraction loss term, $\mathcal{L}_{sim}(\cdot, \cdot)$ is image dissimilarity loss term , and $\mathcal{L}_{seg}(\cdot, \cdot)$ is segmentation loss term.
This equation incorporates bidirectional supervision ($\mathcal{L}_{cls}(\cdot, \cdot)$ and $\mathcal{L}_{ext}(\cdot, \cdot)$), which envelops the entire network to ensure positive forward propagation and controllable feedback across tasks. Additionally, unsupervised and one-shot guidance ($\mathcal{L}_{sim}(\cdot, \cdot)$ and $\mathcal{L}_{seg}(\cdot, \cdot)$) within the model reduces reliance on high-cost annotations. 
The loss terms are scaled by $\alpha$, $\beta$, and $\gamma$ to balance their impacts. 

By leveraging the differentiability in each component of this design, our model achieves joint optimization in an end-to-end manner. All tasks are unified within a single model for collective learning, mutually boosting their performance with limited labels.

%% file: tab_res_ADHD.tex
\begin{table*}[t]
    \centering
    \caption{Results on ADHD dataset. The results are reported as (mean $\pm$ std ) of each task for each compared method.
    “$+$” indicates combining different baselines for the corresponding tasks.
    }
    \label{tab:res ADHD}
    \vspace{-7pt}
    \resizebox{1.0\linewidth}{!}{
    \begin{tabular}{ccccccccccc}
    \toprule
    % \hline

    \multicolumn{1}{c}{\multirow{2}{*}{Methods}}              & 
    \multicolumn{2}{c}{Extraction}             &
    \multicolumn{2}{c}{Registration}    &
    \multicolumn{2}{c}{Segmentation}    &
    \multicolumn{2}{c}{Parcellation}    &
    \multicolumn{2}{c}{Classification}
    
    \\   
    
    % \cmidrule(lr){1-6}
    \cmidrule(lr){2-3}
    \cmidrule(lr){4-5}
    \cmidrule(lr){6-7}
    \cmidrule(lr){8-9}
    \cmidrule(lr){10-11}
    % \hline

    % {}                                        & 
    % {}             &
    % {}             &
    % {}        &
    % {}        &
    {}         &
    
    {Dice $\uparrow$} & 
    {Jaccard $\uparrow$} & 
    
    {MI $\uparrow$} &
    {CC $\uparrow$} &

    {Dice $\uparrow$} &
    {Jaccard $\uparrow$} &

    {Dice $\uparrow$} &
    {Jaccard $\uparrow$} &

    {ACC $\uparrow$} &
    {AUC-ROC $\uparrow$}

    \\
    
    \midrule
    % \hline
    
    BET +
    FLIRT + 
    DW +
    KNN +
    GCN  &
    0.830 $\pm$ 0.058 &
    0.713 $\pm$ 0.079 &
    0.585 $\pm$ 0.031 &
    0.882 $\pm$ 0.041 &
    0.431 $\pm$ 0.058 &
    0.293 $\pm$ 0.049 &
    0.510 $\pm$ 0.172 &
    0.375 $\pm$ 0.142 &
    0.582 $\pm$ 0.034 &
    0.546 $\pm$ 0.028 

    \\   

    % \cmidrule(lr){1-6}
    % \hline
    
    Synth +
    FLIRT + 
    DW +
    KNN +
    GCN  &
    0.920 $\pm$ 0.012 &
    0.853 $\pm$ 0.021 &
    0.621 $\pm$ 0.018 &
    0.942 $\pm$ 0.006 &
    0.494 $\pm$ 0.015 &
    0.347 $\pm$ 0.013 &
    0.678 $\pm$ 0.040 &
    0.525 $\pm$ 0.040 &
    0.595 $\pm$ 0.043 &
    0.612 $\pm$ 0.024 

    \\

    %\midrule
    % \cmidrule(lr){1-6}
    % \hline

    BET + 
    VM + 
    DW + 
    KNN +
    GCN &
    0.830 $\pm$ 0.058 &
    0.713 $\pm$ 0.079 &
    0.584 $\pm$ 0.037 &
    0.874 $\pm$ 0.043 &
    0.432 $\pm$ 0.029 &
    0.296 $\pm$ 0.026 &
    0.599 $\pm$ 0.070 &
    0.442 $\pm$ 0.066 &
    0.578 $\pm$ 0.027 &
    0.568 $\pm$ 0.016 
    
    \\

    % \cmidrule(lr){1-6}
    % \hline
    
    Synth + 
    VM + 
    DW +  
    KNN +
    GCN  &
    0.920 $\pm$ 0.012 &
    0.853 $\pm$ 0.021 &
    0.632 $\pm$ 0.020 &
    0.940 $\pm$ 0.007 &
    0.447 $\pm$ 0.014 &
    0.309 $\pm$ 0.013 &
    0.619 $\pm$ 0.041 &
    0.463 $\pm$ 0.039 &
    0.582 $\pm$ 0.055 &
    0.598 $\pm$ 0.015 

    \\
    
    %\midrule
    % \cmidrule(lr){1-6}
    % \hline
    
    BET + 
    ABN + 
    DW +  
    KNN +
    GCN &
    0.830 $\pm$ 0.058 &
    0.713 $\pm$ 0.079 &
    0.585 $\pm$ 0.036 &
    0.877 $\pm$ 0.043 &
    0.446 $\pm$ 0.031 &
    0.308 $\pm$ 0.027 &
    0.653 $\pm$ 0.051 &
    0.497 $\pm$ 0.051 &
    0.526 $\pm$ 0.036 &
    0.571 $\pm$ 0.017 
    
    \\
    % \cmidrule(lr){1-6}
    % \hline
    
    Synth + 
    ABN + 
    DW +  
    KNN +
    GCN &
    0.920 $\pm$ 0.012 &
    0.853 $\pm$ 0.021 &
    0.635 $\pm$ 0.021 &
    0.943 $\pm$ 0.009 &
    0.455 $\pm$ 0.015 &
    0.317 $\pm$ 0.013 &
    0.675 $\pm$ 0.026 &
    0.521 $\pm$ 0.027 &
    0.595 $\pm$ 0.039 &
    0.612 $\pm$ 0.012 

    \\

    %\midrule
    % \cmidrule(lr){1-6}
    % \hline
    
    ERNet + 
    DW + 
    KNN +
    GCN &
    
    0.935 $\pm$ 0.016 &
    0.879 $\pm$ 0.028 &
    0.636 $\pm$ 0.014 &
    0.952 $\pm$ 0.009 &
    0.498 $\pm$ 0.014 &
    0.350 $\pm$ 0.014 &
    0.677 $\pm$ 0.045 &
    0.523 $\pm$ 0.047 &
    0.582 $\pm$ 0.070 &
    0.612 $\pm$ 0.015 

    \\

    %\midrule
    % \cmidrule(lr){1-6}
    % \hline
    
    BET + 
    DeepAtlas + 
    DW + 
    KNN +
    GCN &
    
    0.830 $\pm$ 0.058 &
    0.713 $\pm$ 0.079 &
    0.587 $\pm$ 0.037 &
    0.874 $\pm$ 0.041 &
    0.478 $\pm$ 0.029 &
    0.344 $\pm$ 0.028 &
    0.591 $\pm$ 0.069 &
    0.434 $\pm$ 0.065 &
    0.599 $\pm$ 0.017 &
    0.579 $\pm$ 0.013

    \\
    % \cmidrule(lr){1-6}
    % \hline
    
    Synth + 
    DeepAtlas + 
    DW + 
    KNN +
    GCN &
    
    0.920 $\pm$ 0.012 &
    0.853 $\pm$ 0.021 &
    0.632 $\pm$ 0.021 &
    0.940 $\pm$ 0.007 &
    0.480 $\pm$ 0.016 &
    0.348 $\pm$ 0.015 &
    0.654 $\pm$ 0.030 &
    0.497 $\pm$ 0.031 &
    0.621 $\pm$ 0.047 &
    0.647 $\pm$ 0.012 

    \\

    %\midrule
    % \cmidrule(lr){1-6}
    % \hline
    
    JERS +
    DW +
    KNN +
    GCN &
    
    0.938 $\pm$ 0.014 &
    0.883 $\pm$ 0.025 &
    0.637 $\pm$ 0.014 &
    0.952 $\pm$ 0.009 &
    0.504 $\pm$ 0.013 &
    0.369 $\pm$ 0.013 &
    0.681 $\pm$ 0.043 &
    0.527 $\pm$ 0.045 &
    0.626 $\pm$ 0.039 &
    0.584 $\pm$ 0.009 

    \\
    
    % \cmidrule(lr){1-6}
    % \hline
    
    JERS +
    DW +
    KNN +
    BGN  &
    
    0.938 $\pm$ 0.014 &
    0.883 $\pm$ 0.025 &
    0.637 $\pm$ 0.014 &
    0.952 $\pm$ 0.009 &
    0.504 $\pm$ 0.013 &
    0.369 $\pm$ 0.013 &
    0.681 $\pm$ 0.043 &
    0.527 $\pm$ 0.045 &
    0.548 $\pm$ 0.085 &
    0.582 $\pm$ 0.094 

    \\

    % \cmidrule(lr){1-6}
    % \hline
    
    JERS +
    DW +
    KNN +
    BNT &

    0.938 $\pm$ 0.014 &
    0.883 $\pm$ 0.025 &
    0.637 $\pm$ 0.014 &
    0.952 $\pm$ 0.009 &
    0.504 $\pm$ 0.013 &
    0.369 $\pm$ 0.013 &
    0.681 $\pm$ 0.043 &
    0.527 $\pm$ 0.045 &
    0.535 $\pm$ 0.039 &
    0.585 $\pm$ 0.034 

    \\

    \midrule
    % \hline
    
    \textbf{UniBrain (ours)} &
    
    \textbf{0.970 $\pm$ 0.003} &
    \textbf{0.942 $\pm$ 0.006} &
    \textbf{0.652 $\pm$ 0.008} &
    \textbf{0.957 $\pm$ 0.008} &
    \textbf{0.520 $\pm$ 0.013} &
    \textbf{0.381 $\pm$ 0.013} &
    \textbf{0.708 $\pm$ 0.019} &
    \textbf{0.557 $\pm$ 0.022} &
    \textbf{0.652 $\pm$ 0.027} &
    \textbf{0.712 $\pm$ 0.030} 

    \\
    
    % \bottomrule   
    \hline
    
    \end{tabular}}
    \vspace{-7pt}
\end{table*}

%% file: tab_methods_v4.tex
\begin{table}[t]
    \centering
    \vspace{-6pt}
    \caption{Summary of compared methods.}
    \label{tab:methods}
    \vspace{-6pt}
    \resizebox{1\linewidth}{!}{
    \begin{tabular}{lcccccc}
    \toprule
    \multirow{2}{*}{\textbf{Methods}}& 
    \multirow{2}{*}{\textbf{Extraction}}& 
    \multirow{2}{*}{\textbf{Registration}}& 
    \multirow{2}{*}{\textbf{Segmentation}}&
    \multirow{2}{*}{\textbf{Parcellation}}&
    \multirow{1}{*}{\textbf{Network}}&
    \multirow{2}{*}{\textbf{Classification}}\\

    {} &
    {} &
    {} &
    {} &
    {} &
    {\textbf{Generation}} &
    {} \\
    
    \midrule
    BET & \cmark  & \xmark & \xmark & \xmark & \xmark & \xmark \\
    
    % BET$^*$ { \cite{smith2002fast}} & \cmark  & \xmark & \cmark & \xmark\\
    SynthStrip & \cmark  & \xmark & \xmark & \xmark & \xmark & \xmark\\

    \midrule
    
    FLIRT & \xmark  & \cmark & \xmark & \xmark & \xmark & \xmark\\
    
    VM & \xmark  & \cmark & \xmark & \xmark & \xmark & \xmark \\

    ABN & \xmark  & \cmark & \xmark & \xmark & \xmark & \xmark \\

    \midrule
    
    DW  & \xmark  & \xmark & \cmark & \cmark & \xmark & \xmark \\

    \midrule
    DeepAtlas & \xmark  & \cmark &
    \cmark & \xmark & \xmark & \xmark\\

    ERNet& \cmark  & \cmark & \xmark & \xmark & \xmark & \xmark\\

    JERS & \cmark & \cmark & \cmark & \xmark & \xmark & \xmark \\

    \midrule
    KNN & \xmark & \xmark & \xmark & \xmark & \cmark & \xmark \\

    \midrule
    %3D-CNN & \xmark  & \xmark & \xmark & \xmark& \xmark & \cmark\\
    
    GCN & \xmark  & \xmark & \xmark & \xmark & \xmark& \cmark\\

    BGN  & \xmark  & \xmark & \xmark & \xmark & \xmark& \cmark\\

    BNT & \xmark  & \xmark & \xmark & \xmark & \xmark& \cmark\\

    \midrule
    UniBrain & \cmark  & \cmark & \cmark & \cmark & \cmark& \cmark\\
    
    \bottomrule
    \end{tabular}
    }
    \vspace{-12pt}
\end{table}

%% file: sec_4_experiment.tex
%\section{Experiment and Analysis}

%\input{tab_res_ABIDE.tex}
\vspace{-5pt}
\section{Experiments}
\label{section:Experiments}

\subsection{Experimental Settings}

\noindent\textbf{Datasets.} We evaluate the effectiveness of our proposed method on the public real-world ADHD dataset with 3D brain sMRI~
\cite{adhd2012adhd}. The dataset contains records for 776 subjects, labeled as real patients (positive) and normal controls (negative). %Out of these, 599 subjects contain ground truth brain extraction and segmentation masks. T
The original dataset is unbalanced, following~\cite{kong2013discriminative}, we randomly sampled 100 ADHD patients and 100 normal controls from the dataset for performance evaluation.
Out of the 200 scans, 160 are used for training, 20 for validation, and 20 for testing. All scans are cropped and resized to $96\times96\times96$ dimensions.
%2) \emph{ABIDE}~\cite{tyszka2014largely} is collected from Autism Brain Imaging Data Exchange dataset. The dataset contains 1112 subjects, labeled as real patients and normal controls. Same to ADHD dataset, we randomly sampled 500 ASD patients and 500 normal controls from the dataset for evaluation. 
We use MNI 152 with the AAL atlas~\cite{tzourio2002automated} as the template image for registration and parcellation. 

\noindent\textbf{Compared Methods.} We compare our UniBrain with several representative baselines. 1) Extraction: BET~\cite{smith2002fast} and SynthStrip~\cite{hoopes2022synthstrip}; 2) Registration: FLIRT~\cite{jenkinson2001global}, VM~\cite{balakrishnan2018unsupervised} and ABN~\cite{su2022abn}; 3) Segmentation and Parcellation: DW~\cite{jaderberg2015spatial}; 4) Network Generation~\cite{zhou2022sparse}; 5) Classification: GCN~\cite{kipf2017semi}, BGN~\cite{li2021braingnn} and BNT~\cite{kan2022brain}; 6) Partial Joint: DeepAtlas (Registration-Segmentation)~\cite{xu2019deepatlas}, ERNet (Extraction-Registration)~\cite{su2022ernet} and JERS (Extraction-Registration-Segmentation)~\cite{su2023one}. Notably, there are no existing solutions that can simultaneously perform all tasks in an end-to-end framework. Thus, for comparison, we designed a pipeline-based solution by combining different state-of-the-art methods for each task. 
The summary of baselines is shown in Table~\ref{tab:methods}.
%and the settings of each baseline are detailed in the Appendix A.6.

%\subsubsection{Experiment Setting.}
\vspace{1pt}
\noindent\textbf{Implementation.} Our experiments are
conducted on Ubuntu 20.04 LTS, utilizing an AMD EPYC
7543 CPU and an NVIDIA Tesla A100-80G GPU.
We split the datasets into training, validation, and test sets as introduced in the Datasets section. The training set is for learning model parameters, the validation set evaluates hyperparameter settings (\eg loss term weights), and the test set is used only once to report the final evaluation results. The code is implemented in Python 3.7.6, and the neural networks are built using PyTorch 1.7.1. The source code is available at~\url{https://github.com/Anonymous7852/UniBrain}.

\input{tab_time_res}

\input{tab_res_ADHD_CNN}

\subsection{Experimental Results}
We compare UniBrain with baseline methods in terms of extraction, registration, segmentation, parcellation, and classification accuracy and efficiency. Additionally, we evaluate UniBrain against voxel-based end-to-end brain imaging analysis solutions, which bypass brain network generation and rely solely on voxel-level information from images for predictions.
Experimental results show that: 1) UniBrain consistently outperforms other methods in extraction, registration, segmentation, parcellation, and classification, while also being time-efficient; 2) UniBrain also surpasses voxel-based end-to-end brain imaging analysis solutions. Similar results were also observed on the ABIDE~\cite{tyszka2014largely} datasets. Due to space constraints, we will present these findings in detail in a future journal publication.
%Additional experiments and evaluation metrics are detailed in Appendix A.2 and Appendix A.3.

%\subsubsection{Overall Results} 
%\label{sec: Overall Results}

%\input{fig_main_res_v2}

\vspace{1pt}
\noindent\textbf{{Overall Results.}} Table~\ref{tab:res ADHD} show the results of the compared methods and the proposed UniBrain in extraction, registration, segmentation, parcellation, and classification tasks. Based on the comprehensive evaluation on the public dataset, UniBrian outperforms existing methods in all tasks. 1) In extraction, we observed that joint-based extraction methods (ERNet, JERS and UniBrain) outperform single-stage extraction methods (BET and Synth). Specifically, UniBrain achieves up to a $5.4\%$ improvement in extraction dice scores over the best single-stage method Synth. 2) For the registration task, methods with strong extraction results typically yield better registration accuracy, highlighting the dependency of accurate registration on prior extraction quality. 3) Good registration enhances segmentation and parcellation performance, as these tasks rely on accurate registration. 4) Classification task results also reflect this trend, with higher parcellation accuracy (like Synth-based, JERS-based, UniBrain) yielding better outcomes due to the classification network leveraging parcellation masks for brain network construction. 
Overall, there's a clear interdependence among brain imaging analysis tasks, with strengths and errors propagating across them. Partially joint methods like ERNet, JERS, and DeepAtlas show improved performance in their joint tasks but are limited when combined with other separate models. In contrast, UniBrain, benefiting from full end-to-end joint learning, uniquely excels across all tasks.

\vspace{1pt}
\noindent\textbf{Running Efficiency.} We measure the efficiency of UniBrain by comparing its inference time with other baselines. The measurement is made on the same device with an AMD EPYC 7543 CPU and an NVIDIA Tesla A100 GPU.
The running time is reported as the average processing time for each image in its corresponding
task.
As indicated in Table~\ref{tab: time ADHD}, fully separate methods are the slowest due to the need for individual optimization of each task. Partially joint learning methods demonstrate increased speed in their joined tasks but still require combination with other methods, limiting overall time efficiency.
UniBrain is the fastest method, which efficiently performs all tasks in an end-to-end manner on the same device, enhancing overall speed.

\vspace{1pt}
\noindent\textbf{Voxel-based End-to-End Learning.} We compare UniBrain with voxel-based end-to-end brain imaging analysis solution. In this experiments, we disregard graph-based models, relying only on voxel information from images for final classification predictions. 
We devised three groups: 1) Direct use of raw MRI images as input (including non-brain tissues, images in different coordinate spaces) for label classification. 2) Use of extracted brain images as input (still in different coordinate spaces) for label classification. 3) Use of the brain been extracted and registered to a standard space as input for classification. As shown in Table~\ref{tab:res ADHD 3D-CNN}, we observed that the performance is worse when using raw images as input due to the inclusion of non-brain tissues and spatial transformation noise. Images processed through extraction and registration yielded higher accuracy. UniBrain, integrating preprocessing and classification in a joint learning approach, outperformed all other models.

%% file: tab_time_res.tex
\begin{table}[t]
    \vspace{-5pt}
    \centering
    \caption{Running time of compared methods on ADHD dataset.}
    \label{tab: time ADHD}
    \vspace{-5pt}
    \resizebox{1.0\linewidth}{!}{
    \begin{tabular}{ccccccc}
    \toprule
    % \hline
    
    \multicolumn{1}{c}{\multirow{2}{*}{Methods}}    & 
    \multicolumn{6}{c}{Time (Sec) $\downarrow$} 
    
    \\
    
    % \cmidrule(lr){1-6}
    \cmidrule(lr){2-7}
    % \hline

    % Ext & Reg  & Seg & Parc & NG & Cls  &
     &
    Ext & Reg  & Seg & Parc & NG & Cls\\

    \midrule
    % \hline
    
    BET + 
    FLIRT + 
    DW +  
    KNN +
    GCN &
    % 1.2452 &
    1.2 &
    % 4.0567 &
    4.1 &
    $1.2 \times 10^{-1}$ &
    $1.5 \times 10^{-1}$ &
    $8.2 \times 10^{-1}$ &
    $2.0 \times 10^{-5}$
    
    \\
    
    % \cmidrule(lr){1-6}\cmidrule(lr){7-12}
    % \hline
    Synth + 
    FLIRT + 
    DW +  
    KNN +
    GCN  &
    % 9.7692 &
    9.8 &
    % 5.2204 &
    5.2 &
    $1.3 \times 10^{-1}$ &
    $1.6 \times 10^{-1}$ &
    $7.9 \times 10^{-1}$ &
    $3.6 \times 10^{-5}$
    
    \\

    % \cmidrule(lr){1-6}\cmidrule(lr){7-12}
    % \hline

    BET +
    VM + 
    DW + 
    KNN +
    GCN  &
    % 1.2452 &
    1.2 &
    $5.7 \times 10^{-3}$ &
    $1.0 \times 10^{-4}$ &
    $1.1 \times 10^{-4}$ &
    $7.9 \times 10^{-1}$ &
    $4.0 \times 10^{-5}$ 

    \\

    % \cmidrule(lr){1-6}\cmidrule(lr){7-12}
    % \hline
    Synth + 
    ABN + 
    DW + 
    KNN +
    GCN  &
    % 9.7692 &
    9.8 &
    $9.2 \times 10^{-3}$ &
    $1.0 \times 10^{-4}$ &
    $1.1 \times 10^{-4}$ &
    $8.2 \times 10^{-1}$ &
    $3.4 \times 10^{-5}$ 
    
    \\
    
    % \cmidrule(lr){1-6}\cmidrule(lr){7-12}
    % \hline
    ERNet +  
    DW +  
    KNN +
    GCN  &
    \multicolumn{2}{c}{$4.0 \times 10^{-2}$} &
    $1.0 \times 10^{-4}$ &
    $1.1 \times 10^{-4}$ &
    $8.0 \times 10^{-1}$ &
    $3.1 \times 10^{-5}$
    
    \\

    % \cmidrule(lr){1-6}\cmidrule(lr){7-12}
    % \hline

    Synth + 
    DeepAtlas + 
    DW +  
    KNN +
    GCN  &
    % 9.7692 &
    9.8 &
    \multicolumn{2}{c}{$7.5 \times 10^{-3}$} &
    $1.1 \times 10^{-4}$ &
    $8.1 \times 10^{-1}$ &
    $4.4 \times 10^{-5}$ 

    \\

    % \cmidrule(lr){1-6}\cmidrule(lr){7-12}
    % \hline
    JERS +
    DW +  
    KNN +
    GCN  &
    \multicolumn{3}{c}{$4.9 \times 10^{-2}$} &
    $1.1 \times 10^{-4}$ &
    $8.1 \times 10^{-1}$ &
    $7.4 \times 10^{-5}$ 

    \\

    % \cmidrule(lr){1-6}\cmidrule(lr){7-12}
    % \hline
    JERS +
    DW +  
    KNN +
    BGN &
    \multicolumn{3}{c}{$4.9 \times 10^{-2}$} &
    $1.1 \times 10^{-4}$ &
    $8.1 \times 10^{-1}$ &
    $2.4 \times 10^{-3}$ 

    \\

    % \cmidrule(lr){1-6}\cmidrule(lr){7-12}
    % \hline
    JERS +
    DW + 
    KNN +
    BNT   &
    \multicolumn{3}{c}{$4.9 \times 10^{-2}$} &
    $1.1 \times 10^{-4}$ &
    $8.0 \times 10^{-1}$ &
    $4.2 \times 10^{-4}$ 

    \\

    \midrule
    % \hline

    \textbf{UniBrain (ours)} & 
    \multicolumn{6}{c}{$2.2 \times 10^{-1}$} 
    
    \\

    \bottomrule  
    % \hline
    \end{tabular}
    }
    \vspace{-15pt}
\end{table}

%% file: tab_res_ADHD_CNN.tex
\begin{table*}[t]
    \centering
    \caption{Voxel-based End-to-End Learning on ADHD dataset}
    \label{tab:res ADHD 3D-CNN}
    \vspace{-5pt}
    \resizebox{0.69\linewidth}{!}
    {
    \begin{tabular}{ccccccc}
    \toprule
    % \hline
    
    \multicolumn{1}{c}{\multirow{2}{*}{Methods}}              & 
    \multicolumn{2}{c}{Extraction}             &
    \multicolumn{2}{c}{Registration}           &
    \multicolumn{2}{c}{Classification}
    
    \\   
    
    % \cmidrule(lr){1-3}
    \cmidrule(lr){2-3}
    \cmidrule(lr){4-5}
    \cmidrule(lr){6-7}
    % \hline

     &
    
    {Dice $\uparrow$} & 
    {Jaccard $\uparrow$} & 
    
    {MI $\uparrow$} &
    {CC $\uparrow$} &

    {ACC $\uparrow$} &
    {AUC-ROC $\uparrow$}

    \\
    
    \midrule
    % \hline

    % - &
    % - & 
    3D-CNN &
    - &
    - &
    - &
    - &
    0.539 $\pm$ 0.048 &
    0.623 $\pm$ 0.014

    \\   

    % \cmidrule(lr){1-3}
    % \hline
    
    BET + 
    3D-CNN &
    0.830 $\pm$ 0.058 &
    0.713 $\pm$ 0.079 &
    - &
    - &
    0.539 $\pm$ 0.021 &
    0.587 $\pm$ 0.019

    \\

    %\midrule
    % \cmidrule(lr){1-3}
    % \hline
    
    Synth + 
    3D-CNN & 
    0.920 $\pm$ 0.012 &
    0.853 $\pm$ 0.021 &
    - &
    - &
    0.547 $\pm$ 0.034 &
    0.634 $\pm$ 0.018

    \\

    % \cmidrule(lr){1-3}
    % \hline
    
    ERNet$_{ext}$ + 
    3D-CNN & 
    0.935 $\pm$ 0.016 &
    0.879 $\pm$ 0.028 &
    - &
    - &
    0.582 $\pm$ 0.044 &
    0.656 $\pm$ 0.025 

    \\
    
    % \cmidrule(lr){1-3}
    % \hline
    
    JERS$_{ext}$ + 
    3D-CNN & 
    0.938 $\pm$ 0.014 &
    0.883 $\pm$ 0.025 &
    - &
    - &
    0.573 $\pm$ 0.037 &
    0.638 $\pm$ 0.020

    \\

    % \cmidrule(lr){1-3}
    % \hline
    
    Synth + 
    FLIRT +  
    3D-CNN &
    0.920 $\pm$ 0.012 &
    0.853 $\pm$ 0.021 &
    0.621 $\pm$ 0.018 &
    0.942 $\pm$ 0.006 &
    0.647 $\pm$ 0.056 &
    0.656 $\pm$ 0.051 
 
    \\

    % \cmidrule(lr){1-3}
    % \hline
    
    Synth + 
    VM + 
    3D-CNN &
    0.920 $\pm$ 0.012 &
    0.853 $\pm$ 0.021 &
    0.632 $\pm$ 0.020 &
    0.940 $\pm$ 0.007 &
    0.617 $\pm$ 0.060 &
    0.651 $\pm$ 0.029 

    \\

    % \cmidrule(lr){1-3}
    % \hline
    
    Synth + 
    ABN + 
    3D-CNN &
    0.920 $\pm$ 0.012 &
    0.853 $\pm$ 0.021 &
    0.635 $\pm$ 0.021 &
    0.943 $\pm$ 0.009 &
    0.634 $\pm$ 0.028 &
    0.622 $\pm$ 0.019 

    \\

    % \cmidrule(lr){1-3}
    % \hline
    
    Synth + 
    DeepAtlas + 
    3D-CNN &
    0.920 $\pm$ 0.012 &
    0.853 $\pm$ 0.021 &
    0.632 $\pm$ 0.021 &
    0.940 $\pm$ 0.007 &
    0.645 $\pm$ 0.039 &
    0.642 $\pm$ 0.031 
    
    \\

    % \cmidrule(lr){1-3}
    % \hline
    
    ERNet + 
    3D-CNN & 
    0.935 $\pm$ 0.016 &
    0.879 $\pm$ 0.028 &
    0.636 $\pm$ 0.014 &
    0.952 $\pm$ 0.009 &
    0.573 $\pm$ 0.042 &
    0.570 $\pm$ 0.022 

    \\

    % \cmidrule(lr){1-3}
    % \hline
    
    JERS +
    3D-CNN & 
    0.938 $\pm$ 0.014 &
    0.883 $\pm$ 0.025 &
    0.637 $\pm$ 0.014 &
    0.952 $\pm$ 0.009 &
    0.613 $\pm$ 0.049 &
    0.596 $\pm$ 0.025

    \\

    \midrule
    % \hline
    
    \textbf{UniBrain (ours)} &
    
    \textbf{0.970 $\pm$ 0.003} &
    \textbf{0.942 $\pm$ 0.006} &
    \textbf{0.652 $\pm$ 0.008} &
    \textbf{0.957 $\pm$ 0.008} &
    \textbf{0.652 $\pm$ 0.027} &
    \textbf{0.712 $\pm$ 0.030}

    \\
    
    \bottomrule  
    % \hline
    
    \end{tabular}}
    \vspace{-10pt}
\end{table*}

%% file: sec_6_conclusion.tex
\vspace{-5pt}
\section{Conclusion}
This paper presents a novel unified framework, UniBrain, the first end-to-end model to jointly perform a diverse set of brain imaging analysis tasks, including extraction, registration, segmentation, parcellation, network generation and classification. UniBrain integrates heterogeneous information into a single system, enabling efficient knowledge transfer across different modules, and avoiding the need for extensive task-specific labels. Experimental results show that UniBrain outperforms state-of-the-art methods in all tasks while also demonstrating robustness and time efficiency.